\documentclass[showpacs,prd,floatfix]{revtex4}

\usepackage{graphicx}
\usepackage{dcolumn}
\usepackage{lineno}
\usepackage{epstopdf}
\usepackage{ifthen}

\newboolean{PRDdraft}
\newboolean{CDFnote}

\setboolean{PRDdraft}{true}  
\setboolean{CDFnote}{false}

\begin{document} 
\bibliographystyle{revtex}
\ifthenelse{\boolean{CDFnote}}{
  \pagewiselinenumbers
   \doublespace

   \hfill CDF/PUB/EXOTIC/CDFR/7764

   \vspace*{1.5cm}

\begin{center}
\begin{Large}
Search for Large Extra Dimensions Using Dielectron and Diphoton
Events in $p\bar{p}$ Collisions at $\sqrt{s}=1.8$~TeV
\end{Large}
\end{center}

\vspace{1in}
\begin{center}
\begin{large}{The CDF Run I Collaboration\\}
\end{large}
\today
\end{center}
\vspace{2.0cm}
}{}

\ifthenelse{\boolean{PRDdraft}}{
\title{Search for Large Extra Dimensions Using Dielectron and Diphoton
Events in $p\bar{p}$ Collisions at $\sqrt{s}=1.8$~TeV}

\author{ David Gerdes$^a$, Simona Murgia$^{c,d}$, John Carlson$^a$, Robert E. Blair$^b$, 
Joey Houston$^c$, Daniel Berebitsky$^a$\\[0.4cm]
{\em $^a$ Department of Physics, University of Michigan, Ann Arbor, MI 48109\\
$^b$ HEP Division, Argonne National Laboratory,  Argonne, IL 60439\\
$^c$ Department of Physics and Astronomy, Michigan State University,  East Lansing, MI  48824\\
$^d$ Department of Physics, Stanford University, Stanford, CA 94305}}

\hspace*{0.2cm}
\date{\today}
}{}

\begin{abstract}
Arkani-Hamed, Dimopoulous, and Dvali have proposed a model of low-scale
quantum gravity featuring large extra dimensions. In this model, 
the exchange of Kaluza-Klein towers of gravitons can enhance the production
rate of electron and photon pairs at high invariant mass in proton-antiproton
collisions. The amount of enhancement is characterized by the parameter
$M_S$, the fundamental Planck scale in the bulk extra dimensions.
We have searched for this effect using 100~pb$^{-1}$ of diphoton data and 110~pb$^{-1}$ of dielectron data collected with the Collider Detector
at Fermilab at $\sqrt{s}=1.8$~TeV during the 1992-1996 run. In the
absence of a signal, we place 95\% confidence-level limits on $M_S$ of $905$ 
GeV/$c^2$ and $826$ GeV/$c^2$, for the case of constructive and destructive
graviton interference respectively.
\end{abstract}

\ifthenelse{\boolean{PRDdraft}}{

\maketitle
}{}

\ifthenelse{\boolean{CDFnote}}{
\newpage
}{}

Theories of low-scale quantum gravity featuring large extra spatial dimensions
(LED) have attracted considerable interest because of their possible
observable consequences at existing and future colliders. In one such scenario,
proposed by Arkani-Hamed, Dimopoulos, and Dvali~\cite{ADD}, the fermions and gauge bosons
of the Standard Model (SM) are confined to the three ordinary spatial dimensions,
which form the boundary (``the brane'') of a space with $n$ compact spatial dimensions
(``the bulk'') in which gravitons alone can propagate. In this model, the Planck scale 
is lowered to the electroweak scale of ${\mathcal{O}}$(1 TeV),  which 
is postulated to be 
the only fundamental scale in nature. 
The fundamental Planck scale in the extra dimensions ($M_S$), the 
characteristic size of the $n$ extra dimensions ($R$) and the Planck scale on
the brane ($M_{Pl}$) are related via

\begin{eqnarray}
M^2_{Pl} \propto M^{n+2}_S R^n,
\label{eqn:led1}
\end{eqnarray}

a purely classical relationship calculated by applying the $4+n$ dimensional Gauss's law. 
In this scenario, then, the weakness of gravity 
compared to the other SM interactions is explained by the suppression 
of the gravitational field flux by a factor proportional to the volume of the 
extra dimensions.

One important consequence for physics in the brane is that the discrete 
momentum modes of excitation of the graviton transverse to the brane propagate 
in our three ordinary dimensions as different mass states.
Analogously to the momentum states, the spacing between these mass states is 
proportional to $1/R$. This collection of mass states forms what is known as a
Kaluza-Klein (KK) tower of gravitons.  The tower can in principle extend up to infinity, 
but there is a cutoff 
imposed by the fact the effective theory breaks down at 
scales above $M_S$. 

The existence of KK gravitons can be tested at colliders by searching for two 
different processes: real graviton emission and virtual graviton exchange.
At leading order, virtual graviton exchange includes processes in which a virtual 
graviton is produced by the annihilation of two SM particles in the initial state, 
the graviton then propagates in the extra dimension and  finally decays into SM particles 
that appear in the brane.
Real graviton production occurs when a graviton is produced together with something 
else by the interaction of SM particles and  escapes into the extra dimensions, leaving 
behind missing energy.

\begin{figure}
\includegraphics[height=2.5cm]{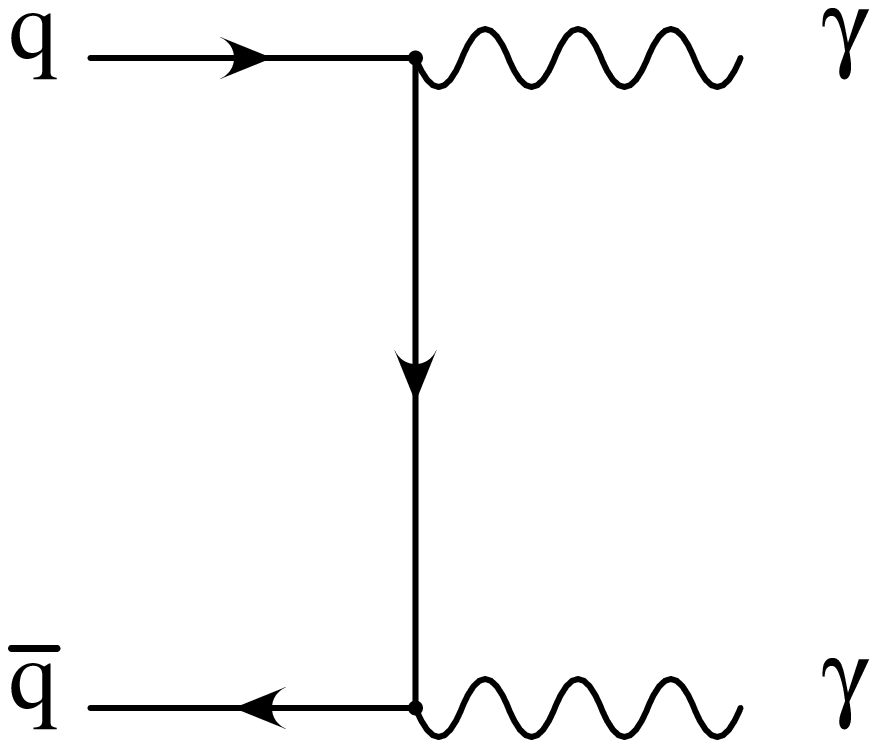}
\includegraphics[height=2.5cm]{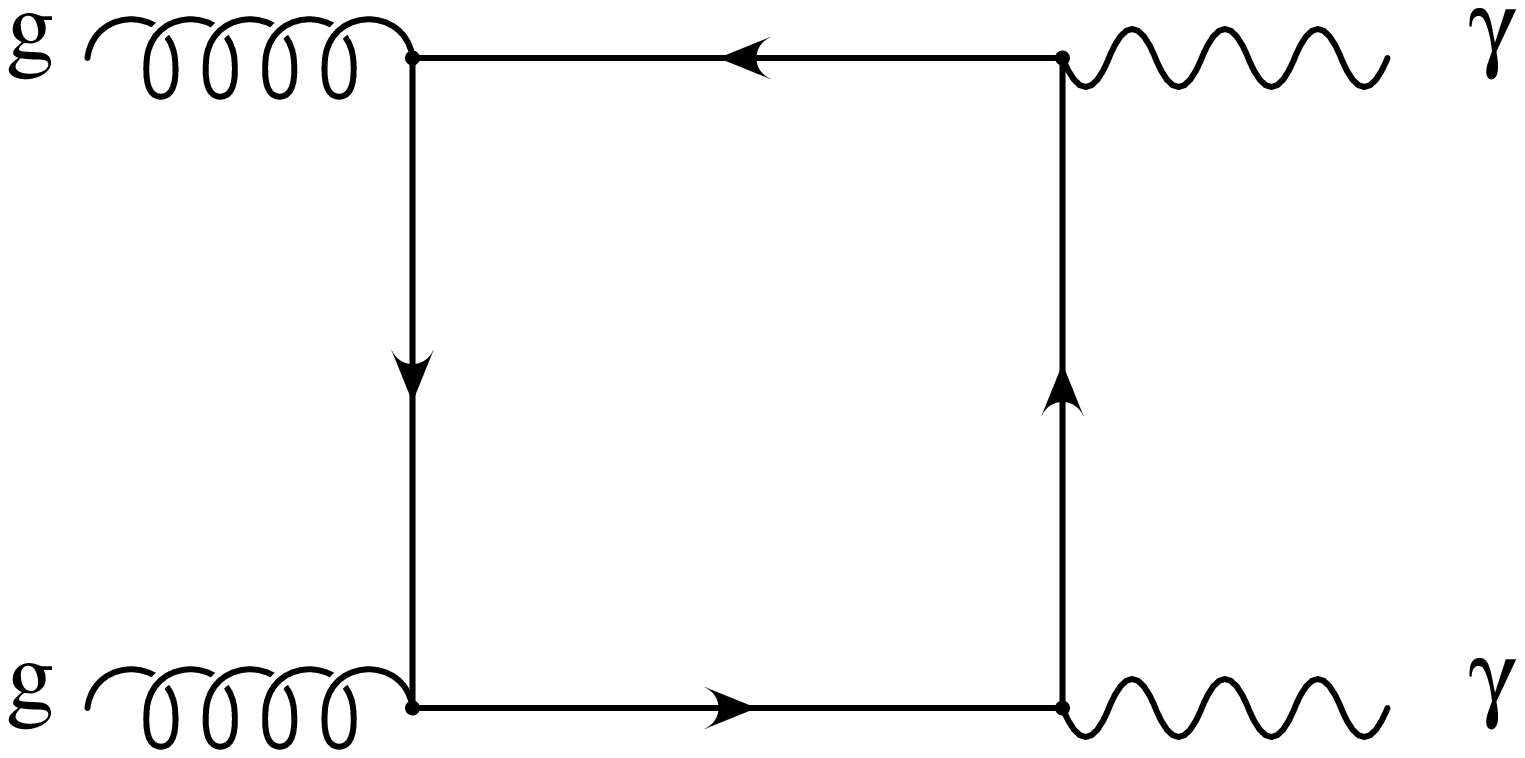}
\includegraphics[height=2.5cm]{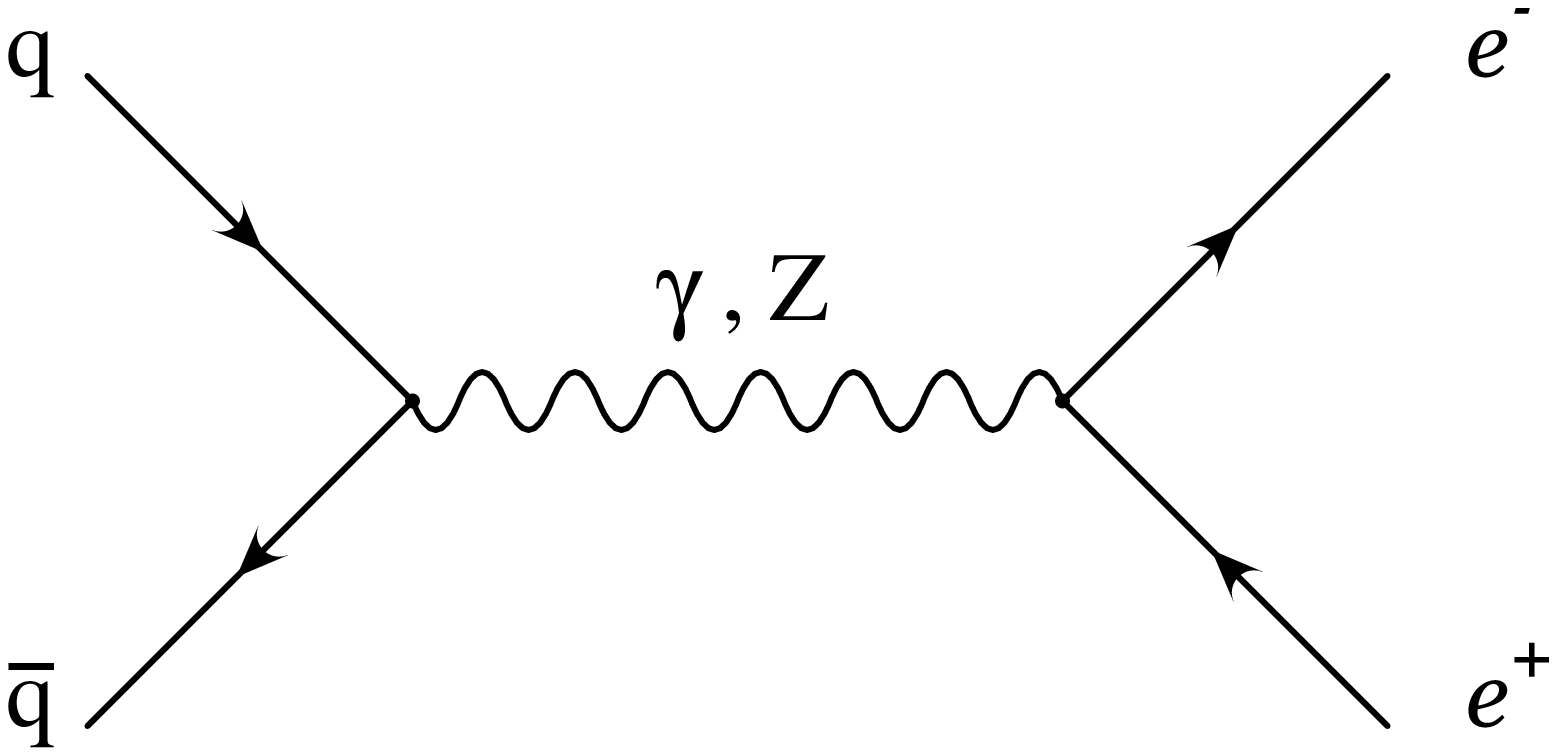}
\caption{Feynman diagrams for SM diphoton and dielectron production at leading order. }
\label{fig:smgg}
\end{figure}

\begin{figure}
\includegraphics[height=2.5cm]{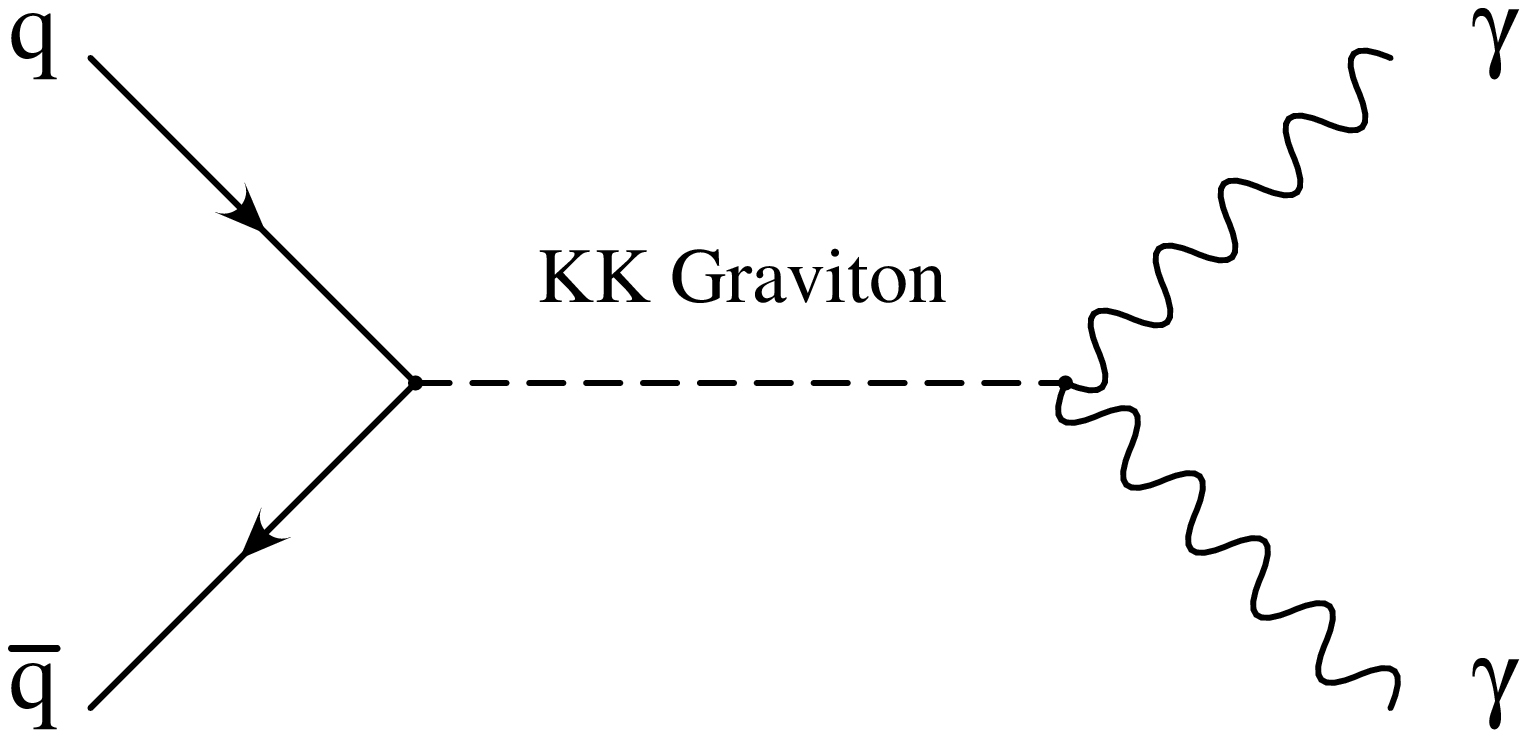}
\includegraphics[height=2.5cm]{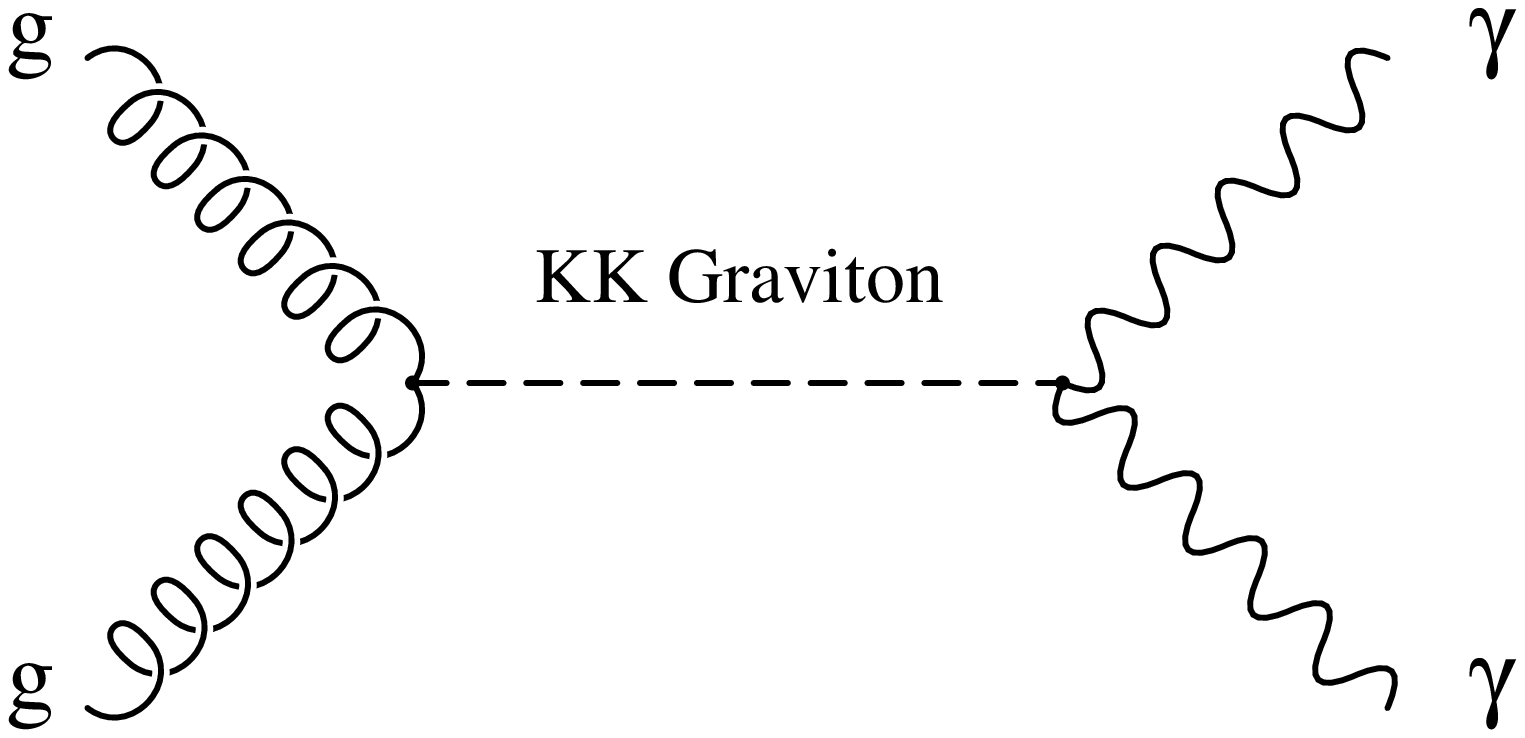}
\\[.1cm]		       
\includegraphics[height=2.5cm]{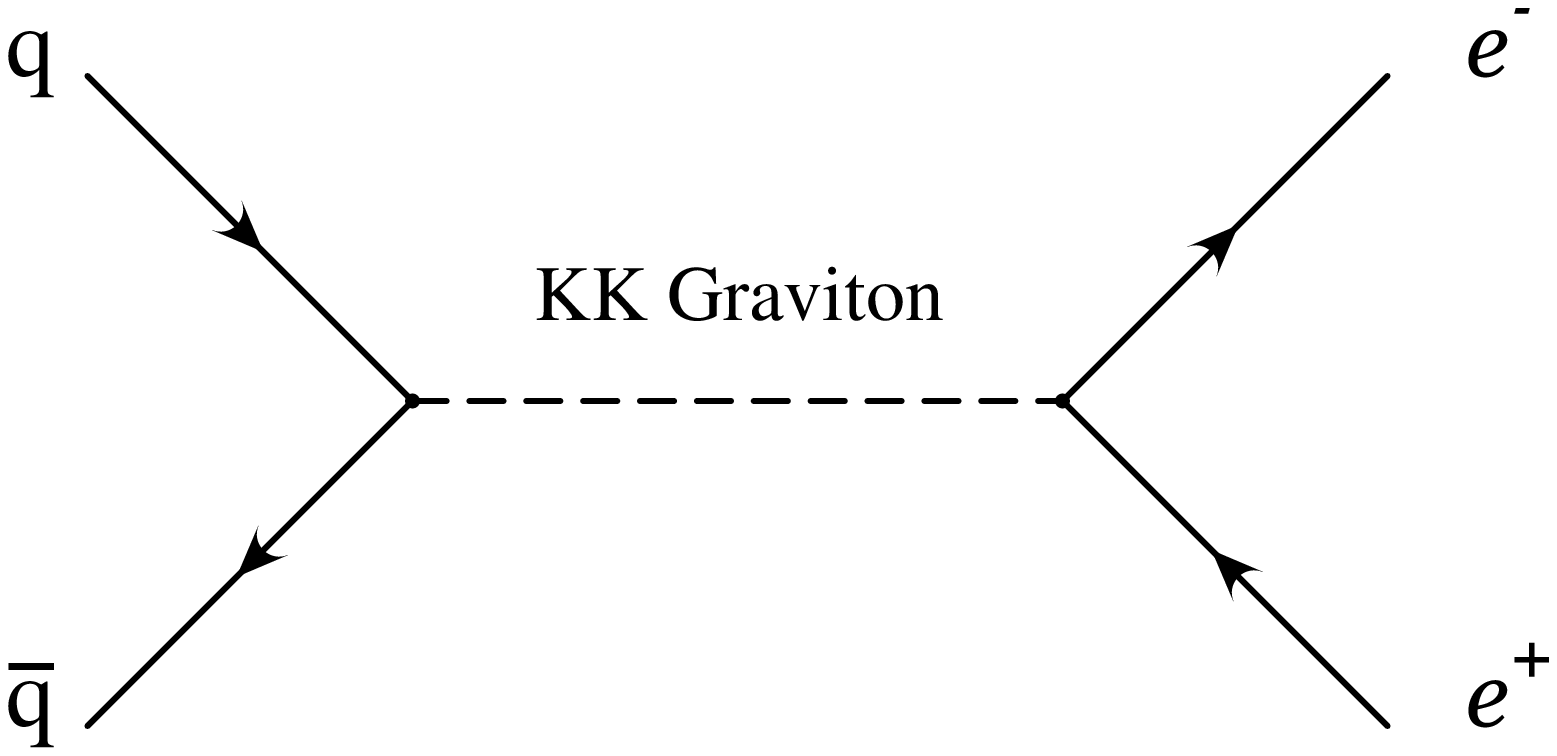}
\includegraphics[height=2.5cm]{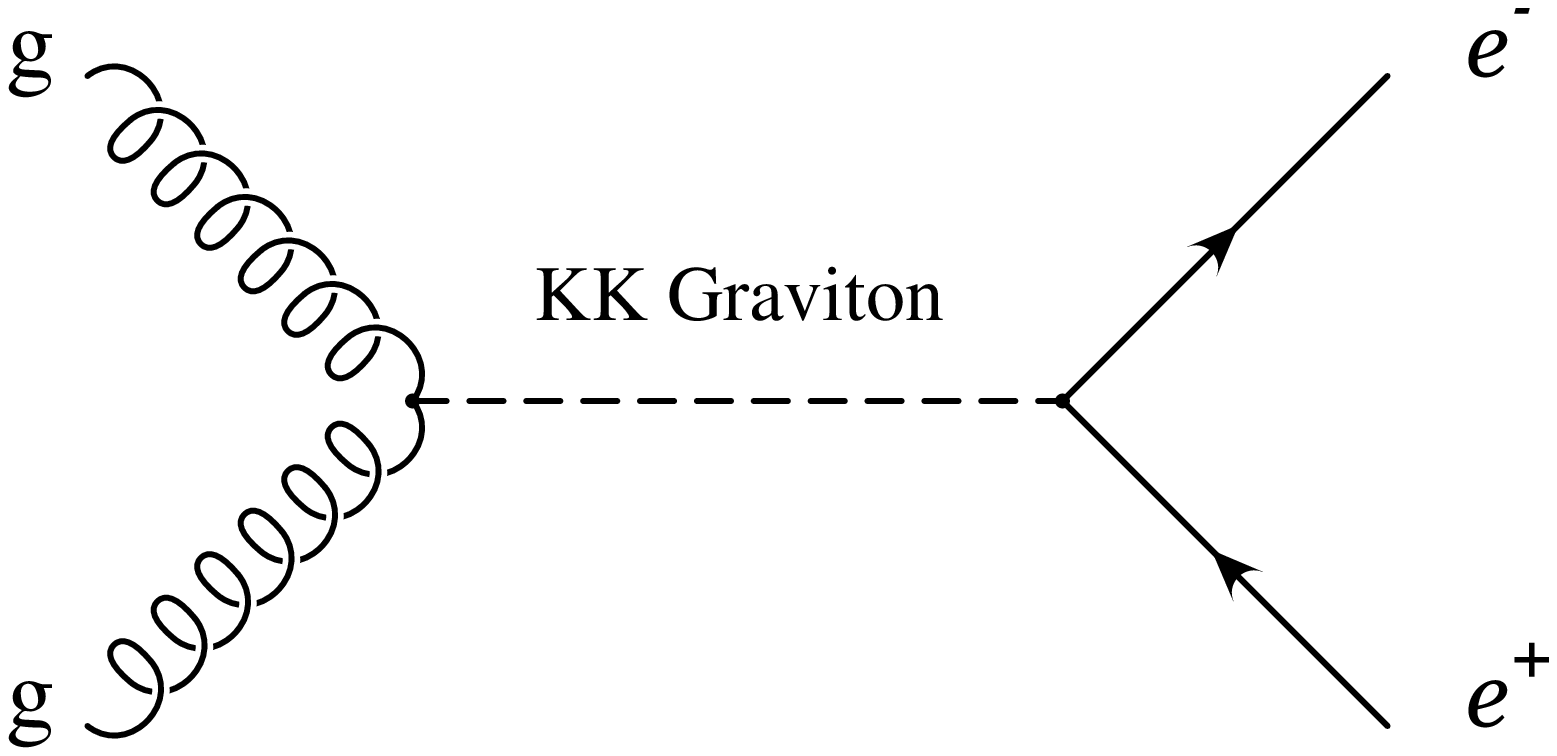}
\caption{Feynman diagrams for LED diphoton and dielectron production at leading order. }
\label{fig:ledgg}
\end{figure}

In this paper, we search for LED by 
looking for the decay of KK gravitons into diphotons~\cite{simona-thesis} or 
dielectrons~\cite{Carlson-thesis}. An analysis of these
final states has previously been carried out by the D\O~Collaboration~\cite{D0prl}.
These channels are desirable because SM production of 
these pairs involves an electromagnetic coupling to the final state;
thus production is suppressed with respect to dijet final
states. However, gravitons couple democratically to stress-energy.
Thus, LED production of dielectron and diphoton final states can
more easily be distinguished from SM sources.
The dominant Feynman diagrams that contribute to these 
processes in the SM are shown in Figure~\ref{fig:smgg}. When KK
gravitons are included, new diagrams appear as shown in Figure~\ref{fig:ledgg}.
Since the LED contribution to 
SM pair production proceeds through a KK tower of graviton states with a closely 
spaced mass spectrum, the extra-dimensional signal
does not appear as a single resonance, but rather as an enhancement of the production 
cross section at high invariant mass where the SM contribution is rapidly 
falling and a large number of gravitons can be produced or, equivalently, more modes
 of the momentum  in the bulk can be excited.

The cross section for graviton production at leading order, as a function of the
 invariant mass of the electron or photon pair, takes the general form~\cite{Cheung-Landsberg}:
 
\begin{equation}
\label{eqn:gamgamxsec2}
\frac{d\sigma}{dM}=  f_{SM}(M)+ \eta_{LED} f_{INT}(M)+ \eta_{LED}^2 f_{KK}(M),
\end{equation}

where $M$ is the invariant mass of the photon or electron pair and
$f_{SM}$, $f_{INT}$, $f_{KK}$ are  terms in the cross section with the
dimensionful parameter $\eta_{LED}$ factored out.
The term $f_{SM}$ on the right-hand side of Equation~\ref{eqn:gamgamxsec2} is the  
contribution from SM processes only. The second term, $f_{INT}$, proportional to the 
first power of $\eta_{LED}$, is the result of the interference between the SM and 
the LED graphs.
Finally, the last term, $f_{KK}$, proportional to $\eta_{LED}^2$, is the contribution coming
 from the direct KK tower exchange.
Graviton-mediated processes introduce the $\eta_{LED}$-dependence in the equation, 
where
$\eta_{LED}=\lambda/M_S^4$ is the parameter that characterizes LED: 
$M_S$ represents the Planck scale in the bulk and should be of order 1 TeV and 
 $\lambda$ is a dimensionless  parameter of ${\mathcal{O}}$(1) and its value is 
model-dependent. 
 The different signs of $\lambda$ allow for different signs of the interference 
between SM and LED  graphs. 
The shapes of the interference and direct KK cross sections 
are independent of $\eta_{LED}$, which affects only the relative and absolute
 normalization, while the shapes themselves depend only on the parton distribution functions (PDF's) and the 
 kinematics of the process. 

\begin{figure}
 \vspace*{0.5cm}
 \includegraphics[height=12cm]{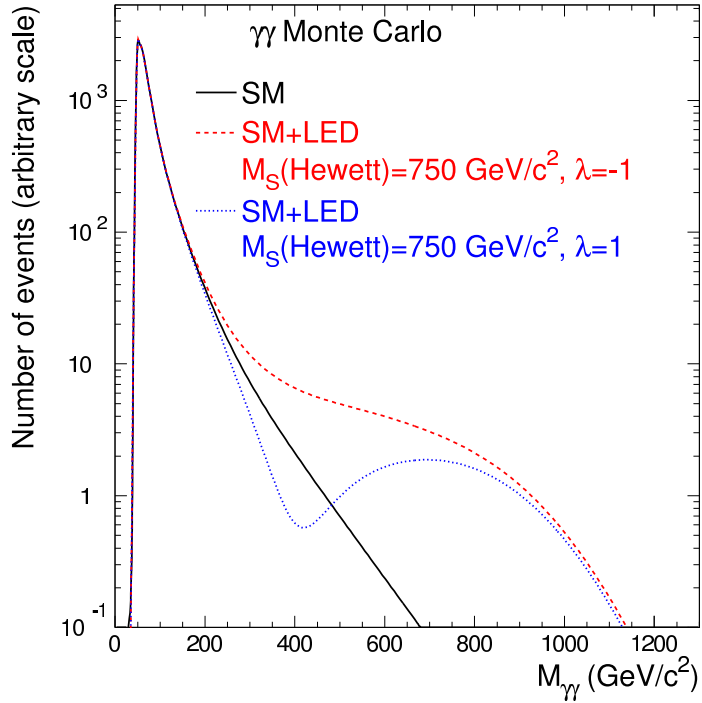}
 \vspace*{-2.5cm}
 \caption{An example of LED ($M_S=750$~GeV/$c^2$) contributions is shown along with the SM prediction in the diphoton invariant mass distribution. The two curves for LED are for constructive ($\lambda=-1$) and destructive ($\lambda=+1$)
interference with the SM processes.}
 \label{fig:MCshapes}
 \end{figure}

As an illustration, Figure~\ref{fig:MCshapes} shows the diphoton invariant mass spectrum 
for the case $M_S=750$~GeV/$c^2$ with constructive interference between the SM and LED
diagrams. The LED signal clearly stands out above the background at higher values
of the invariant mass. The signal in the dielectron case is similar.
We have searched for this signal in approximately 110~pb$^{-1}$ of data collected in
proton-antiproton collisions at $\sqrt{s}=1.8$~TeV by the Collider Detector
at Fermilab (CDF) experiment during the 1992-96 run (Run I). 

The Run I CDF detector has been described in detail in Reference~\cite{CDF}. Here we
present an overview of the detector components most relevant to this 
analysis. The central portion of the detector is cylindrically symmetric around the beam
pipe. Forward/backward extensions together with the central section of the
detector provide close to 4$\pi$ coverage of the solid angle.
Three tracking systems (SVX, VTX, CTC), situated inside a 1.4 T superconducting
solenoid, allow measurements of charged particle momenta for pseudorapidity $|\eta|<1.1$.
In addition, the SVX allows measurement of secondary vertices and the VTX
provides information on the location of the primary vertex along the beam axis.
Located right outside the solenoid coil, the Central Preradiator (CPR), a multi-wire proportional drift chamber, measures tracks of charged particles that result from photon conversions in the inner detector. 
The electromagnetic (EM) and hadronic calorimeter systems are also located outside the 
solenoid coil, past the CPR.
They consist of sampling devices with alternating layers of sensitive  and
absorbing materials, and are segmented in $\eta,\phi$ towers with projective geometry.
The calorimeter is divided
into three components covering different regions of pseudorapidity: 
central ($\mid\eta\mid<$1.1), end-plug
(1.1$<\mid\eta\mid<$2.4) and forward (2.4$<\mid\eta\mid<$4.2).
Only the central and end-plug regions are used in this analysis.
The central strip chambers (CES) are embedded into the central EM calorimeter section, 
and provide a high-resolution measurement of the lateral shower profile in $z$ and $r-\phi$ at 
the point of its maximum development. Finally, muon chambers are located outside the calorimeter to detect particles that escape the inner layers of the detector. 

The diphoton sample consists of two distinct subsets, corresponding 
to two different event topologies.
One of the subsets, the Central-Central (CC) diphoton sample, consists of
events with both photons in the central region of the detector, while the other, 
the Central-Plug (CP) diphoton sample, 
has events with at least one central electromagnetic cluster and at least one 
cluster in the end-plug electromagnetic calorimeter.
The CC diphoton sample has been described in Reference~\cite{ggsearch}.
Photons in this sample are required to have transverse energy ($E_T$) $>22$~GeV, to pass a calorimeter isolation cut, to
have no nearby charged tracks, and to have a CES shower profile consistent with that of a photon. The identification efficiency of the CC diphoton selection is 0.626$\pm$0.017, and the integrated luminosity of this sample is ${\rm 100~pb^{-1}}$. A total of 287 CC events pass these cuts. 

Two trigger paths are used to select the CP diphoton sample.
The first path requires at least one isolated central EM cluster with  
${E_T}>$ 23~GeV, while the second path requires an ${E_T}>$ 50~GeV central EM cluster 
with no calorimeter isolation requirement. The central cluster must pass the the same 
offline selection 
applied to the CC diphoton sample but with an ${E_T}>$ 25~GeV to ensure a high trigger efficiency. 
The plug EM cluster is required to have ${E_T}>$ 22~GeV.
CTC tracking does not extend to the plug region of the detector; thus to
reject charged particles we 
require that the fraction of hit wires in the VTX in a road centered on
the cluster direction is smaller than 0.4.
We require the ${E_T}$ in the annulus between 0.4 and 0.7 in $\eta$-$\phi$
space around the cluster to be no larger than 4 GeV. This requirement allows us
to select isolated clusters while avoiding possible inefficiency
due to additional leakage into the cone immediately surrounding
the cluster that occurs at larger rapidities.
The efficiency of this cut is calculated with electrons from a $Z \rightarrow 
e^+ e^-$ control sample, and is found to be $89\%$, 10\% more than 
the canonical isolation cut based on a cone of 0.4. 
The identification efficiency of the CP diphoton selection is 0.550$\pm$0.065. In our integrated luminosity 
of ${\rm 87~pb^{-1}}$, 192 CP events pass our cuts.

Dielectron events used in this analysis are required to have at least
one ``tight'' electron candidate in the central calorimeter, while the second 
``loose'' candidate can be in either the central or plug calorimeter. 
The details of the selection requirements are described in References~\cite{zprime,Carlson-thesis}.
All events must pass a central electron trigger that requires an EM cluster with  
$E_T>22$~GeV together with a track with $p_T> 13$~GeV/$c$.
Offline, the tight electron must pass an $E_T$
cut of 25 GeV and have the ratio of energy to momentum ($E/p$)$<4$. For tracks with $p_T>50$~GeV/$c$, 
no $E/p$ cut is imposed. We require
good agreement between the EM shower location and the extrapolated
track position, and require the ratio of hadronic to electromagnetic
energy in the cluster to be low. Finally, we require the electron to be
isolated. The loose EM cluster is also required to have
$E_T>25$~GeV, but central candidates may pass more relaxed 
isolation and hadronic energy cuts. Plug electron candidates
are required to have a low hadronic energy fraction, to be 
isolated, and to have a lateral shower profile consistent with
that of testbeam electrons. The efficiency of the electron 
identification cuts is measured using $Z$ data and is found to be 
$0.916\pm 0.010$ for plug electrons and $0.961\pm0.024$ for central electrons.
In an integrated luminosity of $110$ pb$^{-1}$, we identify
3319 CC and 3825 CP candidates. 

Both diphoton and dielectron events are generated using a leading-order 
matrix-element Monte Carlo (MC) program~\cite{Baur}.
The simulation includes the matrix elements 
for direct SM diphoton production with $q\bar{q}$ or 
$gg$ in the initial state, production via KK exchange from a 
$q\bar{q}$ or $gg$ initial state, as well as interference between the SM and 
LED diphoton production processes with $q\bar{q}$ 
in the initial state. However, it does not include the interference between 
the $gg$-initiated SM and LED initial states, which  
is  expected to be small~\cite{Eboli}. A conservative estimate of this 
contribution is incorporated
 into the systematic uncertainties.
The simulation also generates SM $q\bar{q}\rightarrow e^+ e^-$ and
$q\bar{q}$ or $gg\rightarrow e^+ e^-$ events mediated by a KK tower of gravitons.
The \textsc{pythia} MC program~\cite{PYTHIA} is used to model 
fragmentation, parton showers, and the underlying event with CTEQ5M PDF's~\cite{CTEQ5M}.
The generated events are passed through a fast simulation of the CDF detector.
We select simulated events using 
the same cuts applied to the data. 
This procedure yields a prediction of $96\pm31$ prompt SM diphoton events
in the CC sample and $76\pm 14$ events in the CP sample.
 In the dielectron sample, the prediction is $3463\pm223$ CC and $3883\pm292$ CP prompt SM dielectron events. 

However,  the bulk of the diphoton sample consists of events in which one or both 
electromagnetic clusters originate from jets. Hard jet fragmentation to leading
neutral mesons ($\pi^{0}, \eta$) 
that subsequently decay into multiple photons can fake the prompt photon signature. 
In the CC case, the fake background is evaluated by employing statistical
 methods~\cite{CESCPR} that use CES and CPR pulse shape and height information to 
discriminate between prompt photons and photons 
originating from neutral meson decays.
 These techniques yield a fake estimate of 64$\pm$11$\pm$19\%,
 which corresponds to 183$\pm$56$\pm$32 events.
The mass spectrum of the CC fake background is determined from the invariant mass 
distribution of events in a control sample of non-isolated diphotons~\cite{ggsearch}. 

In the CP case, we rely on the above techniques to 
determine the probability that the central cluster is a jet. 
However, these methods cannot be used for plug clusters. 
Instead, we employ dijet and non-isolated diphoton 
 control samples to estimate the fake rate of events in which the central 
cluster is a prompt photon while the plug cluster is a jet~\cite{simona-thesis}.
The total number of fake events in the CP sample is 132$\pm$31. 
The mass spectrum of the CP fake background is determined 
by simulating SM $\gamma$+jet processes using the \textsc{pythia} 
MC program and the CDF detector simulation. These events 
provide the mass spectrum for $\gamma$+fake events. The mass spectrum for 
events where both clusters are fakes is obtained by multiplying the MC $\gamma$+jet
 shape by the ratio between the  $\pi^0 \pi^0$ and $\gamma \pi^0$  cross sections 
calculated at leading order~\cite{owens}.
Because $\pi^0$'s are the main contribution to the fakes, we take this rescaled 
shape as the shape of the fake+fake contribution.

In summary, in the diphoton channel the prediction of fakes plus SM $\gamma\gamma$ is 280$\pm$66  
events in the CC sample and 208$\pm$34 events in the CP sample, in 
good agreement with the observed 287 events in the CC case and 192 events 
in the CP case.

In the dielectron sample, the electrons in the final state are produced by other SM processes in addition to Drell-Yan.
The dominant contribution is from
QCD dijet events in which each jet contains a real or fake electron that 
survives the electron identification cuts.
This mass distribution and size of this background is evaluated using samples 
of non-isolated dielectrons. We obtain 
10$\pm$6 background events in the CC case and 224$\pm$17 background
 events in the CP case.

 In summary, in the dielectron channel the number of events predicted, adding together background and SM contributions, is $3473\pm223$ CC and $4107\pm293$ CP dielectron events, in agreement with the observed 3319 events in the CC case and 3825 in the CP case.

\begin{figure}
\vspace*{1cm}
\includegraphics[height=9cm]{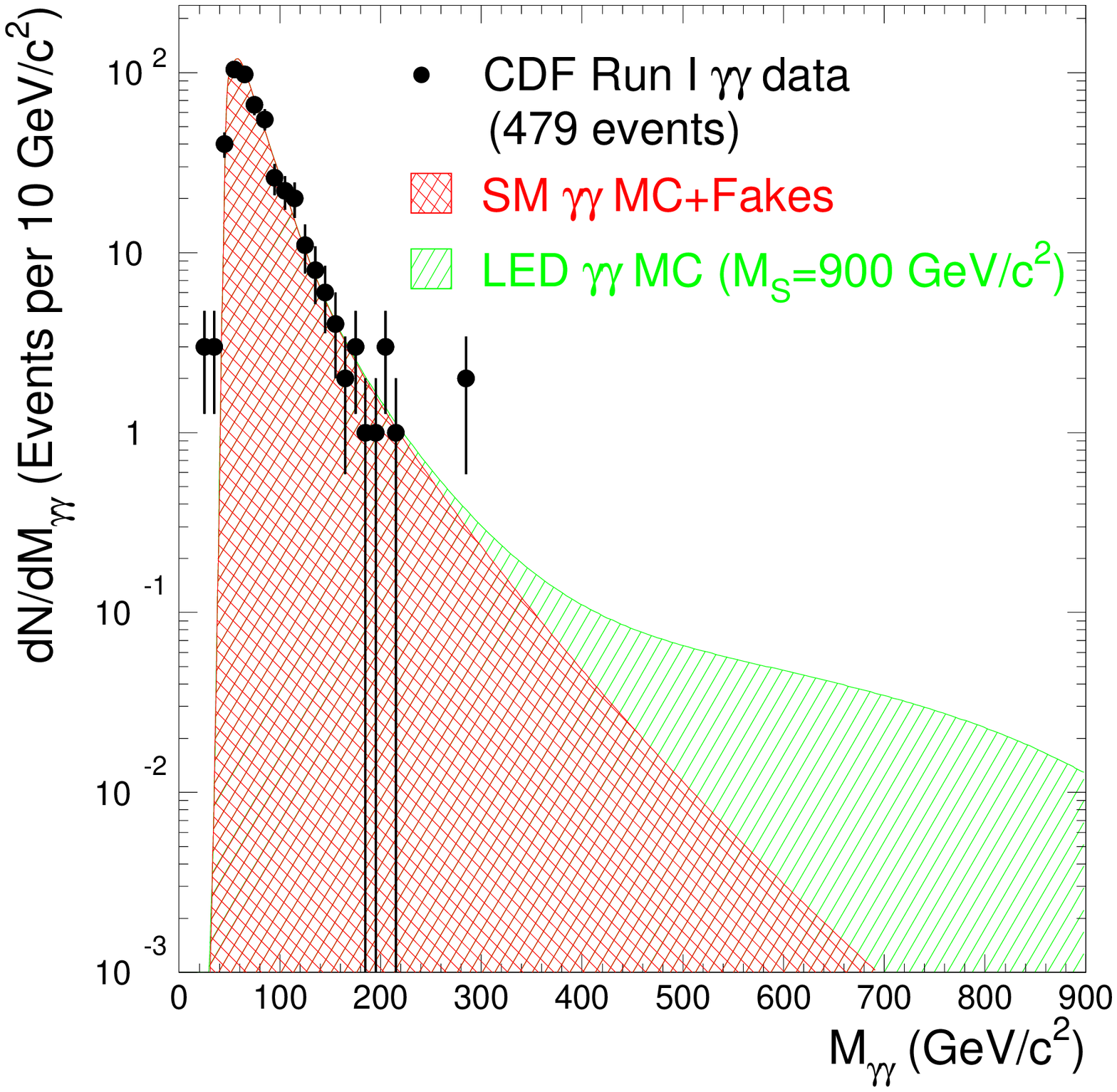}
\\[-0.5cm]
\includegraphics[height=9cm]{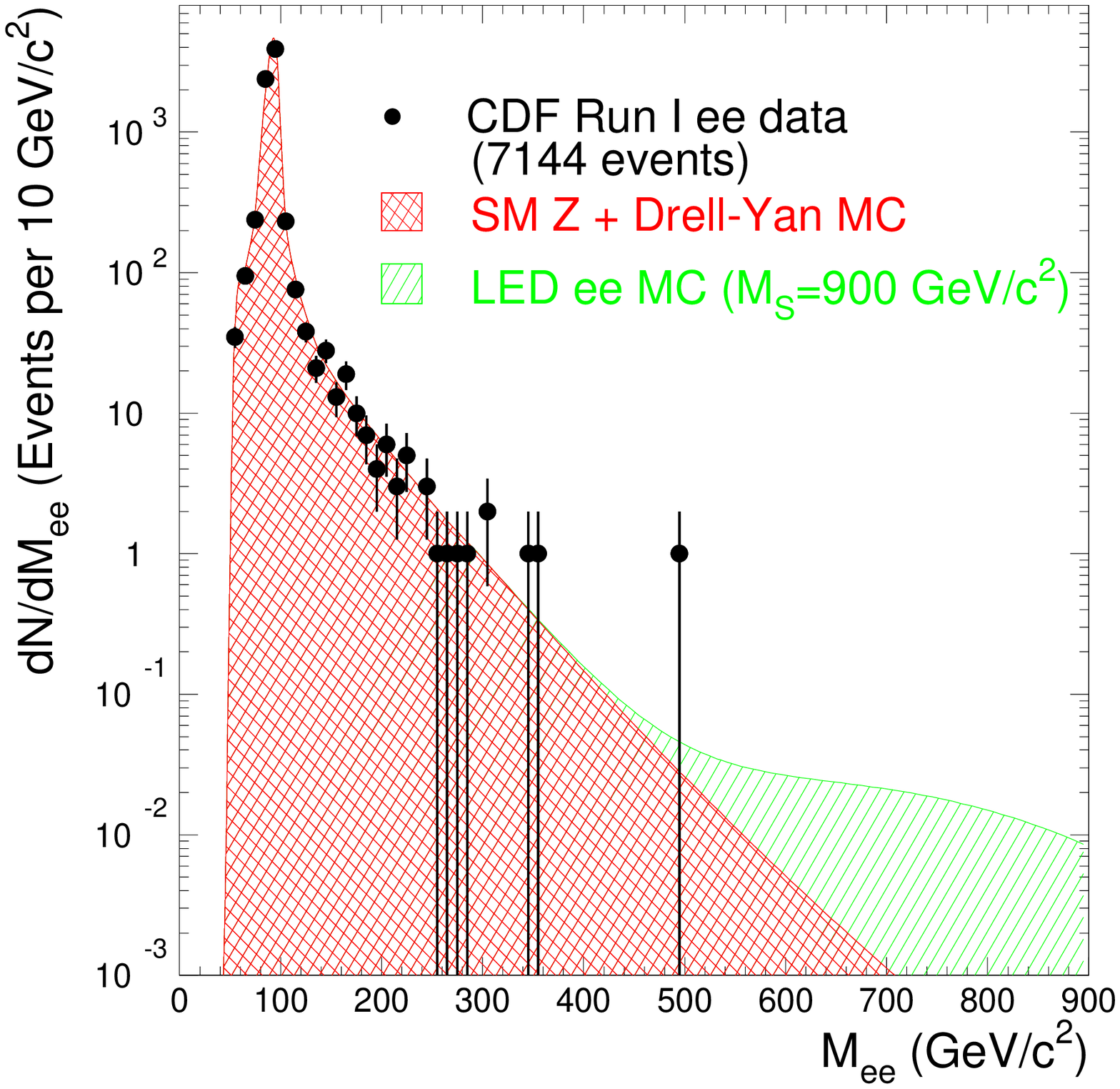}
\vspace*{-2.5cm}
\caption{Invariant mass distributions for the diphoton (top)  and dielectron (bottom) data overlaid with the expected invariant mass distributions from SM and fake sources, as well as the LED contribution for ${ M_S({\rm Hewett})}=900~{\rm GeV}/c^2$ .}
\label{fig:combmass}
\end{figure}

We find that the data are in good agreement with the SM plus fakes prediction both with respect to the rate and invariant mass distribution.
Figure~\ref{fig:combmass} shows the data overlaid with the background 
predictions for diphoton and dielectron events.
We therefore proceed 
to set a limit on LED by determining the lower bound on 
the parameter $M_S$ that characterizes the strength of the graviton coupling. 
We do this by performing a likelihood fit to the diphoton and dielectron
invariant mass distributions.
We use two very similar likelihood functions, one for the diphoton and one
for the dielectron case, which are multiplied together to obtain the 
overall likelihood function ${\cal L}$. The form of ${\cal L}$ is similar to
that employed in CDF's measurement of the top quark mass~\cite{topmass}.
In the likelihood function, a Gaussian constraint
is applied to the number of SM and fake dielectron and diphoton events 
in the CC and CP samples. In addition, a Poisson constraint is applied to the total
number of dielectron and diphoton events. Finally, a shape term weights 
the invariant mass of each event according to the consistency of its invariant mass 
with the SM, background, and LED shapes. We minimize $-\ln {\cal L}$ to obtain
the best-fit value of $\eta_{LED}=\lambda/M_S^4$ for the cases $\lambda=\pm1$, corresponding
to constructive and destructive interference of the SM and KK terms. These
values are consistent with zero, so we extract 95\% confidence level (C.L.) limits. 
For the diphoton data alone, the limits on $\eta_{LED}$ for constructive and 
destructive interference respectively are $\eta_{LED}$=1.36 TeV$^{-4}$ and 
$\eta_{LED}$=2.18 TeV$^{-4}$.
For dielectrons, the limits are  $\eta_{LED}$=2.69 TeV$^{-4}$ (constructive)
 and $\eta_{LED}$=2.86 TeV$^{-4}$ (destructive). 

Systematic uncertainties in this analysis arise from effects that can modify the 
shapes of the SM, fake, and LED invariant mass distributions. In both the electron 
and photon analyses, the dominant
effects are the choice of PDF's, variations in initial state radiation (ISR),
and the shape of the CP fake background. Furthermore, the interference between 
SM and LED $gg$ initial states is not included in the diphoton MC. 
To quantify these systematics, we generate shifted
MC invariant mass distributions with each systematic effect separately 
varied. For example, we vary the choice of PDF's (CTEQ5HJ, MRST(h-g), 
MRST(larger $d/u$ ratio), MRST(low-$\alpha_s$), MRST($q_d$), instead of the default CTEQ5M~\cite{CTEQ5M}), and turn off the effects of ISR. 
To estimate the uncertainty from the neglected $gg$ interference term, we note that
the cross section for the $gg\rightarrow\gamma\gamma$ process is about
$30\%$ of the $q\bar{q}$ process, so the $gg$ interference term should be roughly
$(0.3)^2\approx 0.1$ of the $q\bar{q}$ one.  We vary the $q\bar{q}$ interference 
portion of the templates by twice this amount, $\pm 20\%$.
The systematic uncertainty on the CP background is estimated by using the  
second most populated region of the non-isolated dielectron control sample to determine
the background shape, and computing the change in fitted $\eta_{LED}$ versus the default (most 
populated region).  
For each systematic effect, many pseudo-experiments are 
generated at different values of $M_S$ from the shifted distributions, and the resulting invariant mass distributions are fitted to our default templates. For each systematic
effect, we determine the average shift between the input value of $\eta_{LED}$ and the value 
extracted from the fit. The resulting shifts for all systematic effects are added in quadrature.
The overall shift is used as the width of a gaussian that is convolved with the raw likelihood
function from the fit to the data. Remaining uncertainties on efficiencies, 
acceptance, luminosity, and background normalization are already included in the 
width of the gaussian constraints in the likelihood function.

In the diphoton analysis, the 95\% C.L. limits with systematics included  are  
$\eta_{LED}$=1.53 TeV$^{-4}$ (constructive interference) and $\eta_{LED}$=2.48 TeV$^{-4}$ (destructive). 
In the dielectron case, we find $\eta_{LED}$=2.70 TeV$^{-4}$ (constructive) and 
$\eta_{LED}$=2.87 TeV$^{-4}$ (destructive).

Finally, we obtain the strongest limits by combining the diphoton and dielectron samples. 
A new fit is performed by multiplying the likelihood functions for each sample and minimizing 
it with respect to the common parameter $\eta_{LED}$.
The systematic uncertainties are included into the combined limit in the following way. 
The full systematic uncertainties for the two search channels, $\Delta\eta_{LED,ee}(\eta_{LED})$ 
and $\Delta\eta_{LED,\gamma\gamma}(\eta_{LED})$, are broken into their component pieces, 
$\delta\eta_{LED,ee}^{PDF}$, $\delta\eta_{LED,ee}^{ISR}$, $\delta\eta_{LED,ee}^{uncorr}$, 
$\delta\eta_{LED,\gamma\gamma}^{PDF}$, $\delta\eta_{LED,\gamma\gamma}^{ISR}$ and  
$\delta\eta_{LED,\gamma\gamma}^{uncorr}$. Here $\delta\eta_{LED,ee}^{uncorr}$ represents the
 sum in quadrature of the dielectron systematics with the diphoton systematics, 
i.e. background and LED acceptance. Similarly, $\delta\eta_{LED,\gamma\gamma}^{uncorr}$
 represents the sum in quadrature of the diphoton systematics uncorrelated with the 
dielectron systematics, i.e. SM+LED interference cross section and CP 
background. Each likelihood function, diphoton and dielectron, is smeared by taking 
the corresponding, unsmeared likelihood function, and smearing each point by 
throwing random numbers $\sigma_i$, i=1,...,4, and smearing  a sample of values
 of $\eta_{LED}$ in the following way:
 
\begin{eqnarray}
\eta_{LED,ee}'&=&\eta_{LED,ee}+\sigma_1\delta\eta_{LED,ee}^{PDF}(\eta_{LED,ee})+
\sigma_2\delta\eta_{LED,ee}^{ISR}(\eta_{LED,ee})+\sigma_3\delta\eta_{LED,ee}^{uncorr}(\eta_{LED,ee})
\\
\eta_{LED,\gamma\gamma}'&=&\eta_{LED,\gamma\gamma}+
\sigma_1\delta\eta_{LED,\gamma\gamma}^{PDF}(\eta_{LED,\gamma\gamma})+
\sigma_2\delta\eta_{LED,\gamma\gamma}^{ISR}(\eta_{LED,\gamma\gamma})+
\sigma_4\delta\eta_{LED,\gamma\gamma}^{uncorr}(\eta_{LED,\gamma\gamma}).
\end{eqnarray}

This ensures that the uncorrelated uncertainties are smeared independently while 
the correlated PDF's and ISR uncertainties are smeared together. 

After the smearing procedure is performed, the  diphoton and dielectron smeared 
likelihood functions are multiplied together and the same method described above 
is used to calculate the combined dielectron-diphoton 95\% C.L. limit on $\eta_{LED}$.
We find  $\eta_{LED}$=1.49 TeV$^{-4}$ (constructive) and $\eta_{LED}$=2.15 TeV$^{-4}$ (destructive).

The limits on $\eta_{LED}$ are converted into limits on $ M_S$ by using the 
equation $\eta_{LED}=\lambda/{M^4_S}$. 
In Hewett convention~\cite{hewett}, $\lambda=-1$ for constructive interference and $\lambda=+1$ 
for destructive interference. 
Other popular conventions are those of Giudice, Rattazzi, Wells (GRW)~\cite{GRW}, and  Han, 
Lykken, Zhang (HLZ)~\cite{HLZ}. To translate results from Hewett to GRW convention one simply multiplies 
${ M_S({\rm Hewett})}$ by $\sqrt[4]{\pi/2}$ (constructive interference only).
In HLZ convention the dependence on the number of extra dimensions $n$ is
 calculated and it is incorporated into $\lambda$. For $n>2$:
 
\begin{eqnarray}
{ M_S({\rm HLZ})= \sqrt[4]{\frac{\pi}{2}\left(\frac{2}{2-{\it n}}\right)}\: { M_S({\rm Hewett})}}.
\end{eqnarray}

The dielectron and diphoton combined limits in these conventions are summarized 
in Table ~\ref{tbl:MS95}.
The limits quoted so far assume a $K$-factor of 1.0 for the LED process. 
Limits can be improved somewhat by assuming $K=1.3$ to account for next-to-leading order effects for the SM diphoton and dielectron production as well as for the corresponding graviton mediated processes, as was done in Reference~\cite{D0prl}. 
These limits are also shown in Table~\ref{tbl:MS95}.

\begin{table}
\caption{Diphoton and dielectron limits on ${ M_S}$ in Hewett, GRW, and HLZ conventions. The final
two rows show the combined limits with and without a $K$-factor. \vspace*{5mm}}
\begin{tabular}{|l|l|cc|c|ccccc|}  
 \hline \multicolumn{10}{|c|}{${ M^{95}_S}$ (GeV/$c^2$)} \\ \hline
 &  & \multicolumn{2}{c|} {Hewett} & {GRW} & \multicolumn{5}{c|} {HLZ} \\ 
Sample & $K$ & $\lambda$=-1 & $\lambda$=+1 & & $n$=3 & $n$=4 & $n$=5 & $n$=6 & $n$=7\\\hline\hline
CC $\gamma\gamma$    &     & 870 & 808 &     &      &      &     &     &  \\  
CP $\gamma\gamma$    &     & 718 & 639 &     &      &      &     &     &  \\
CC+CP $\gamma\gamma$ & 1.0 & 899 & 797 & 1006 & 1197 & 1006 & 909 & 846 & 800 \\
\hline
$e^+e^-$ & 1.0 & 780 & 768 & 873 & 1038 & 873 & 789 & 734 & 694 \\ \hline
$e^+e^-$+$\gamma\gamma$ & 1.0 & 905 & 826 & 1013 & 1205 & 1013 & 916 & 852 & 806 \\
$e^+e^-$+$\gamma\gamma$ & 1.3 & 939 & 853 & 1051 & 1250 & 1051 & 950 & 884 & 836 \\ 
\hline \hline 
\end{tabular}
\label{tbl:MS95}
\end{table}

      We thank the Fermilab staff and the technical staffs of the
participating institutions for their vital contributions.  This work was
supported by the U.S. Department of Energy and National Science Foundation;
the Italian Istituto Nazionale di Fisica Nucleare; the Ministry of Education,
Culture, Sports, Science, and Technology of Japan; the Natural Sciences and
Engineering Research Council of Canada; the National Science Council of the
Republic of China; the Swiss National Science Foundation; the A. P. Sloan
Foundation; the Bundesministerium fuer Bildung und Forschung, Germany; the
Korea Science and Engineering Foundation (KoSEF); the Korea Research
Foundation; and the Comision Interministerial de Ciencia y Tecnologia, Spain. 

\end{document}